\documentstyle[aas2pp4]{article}

\begin{document}
\title{A Non-Triggered Burst Supplement to the BATSE Gamma-Ray Burst
  Catalogs}

\author{Jefferson M. Kommers,\altaffilmark{1} Walter H. G.
  Lewin,\altaffilmark{1} Chryssa Kouveliotou,\altaffilmark{2,3} Jan van
  Paradijs,\altaffilmark{4,5} Geoffrey N. Pendleton,\altaffilmark{4}
  Charles A. Meegan,\altaffilmark{3} and Gerald J. Fishman\altaffilmark{3}}

\altaffiltext{1}{Department of Physics and Center for Space Research,
  Massachusetts Institute of Technology, Cambridge, MA 02139; {\tt kommers@space.mit.edu}}
\altaffiltext{2}{Universities Space Research Association, Huntsville, AL 35800}
\altaffiltext{3}{NASA/Marshall Space Flight Center, Huntsville, AL
  35812}
\altaffiltext{4}{University of Alabama in Huntsville, Huntsville, AL
  35812}
\altaffiltext{5}{University of Amsterdam, Amsterdam, Netherlands}


\begin{abstract}
  The Burst and Transient Source Experiment (BATSE) on the {\it
    Compton Gamma Ray Observatory\/} detects gamma-ray bursts (GRBs)
    with a real-time burst detection (or ``trigger'') system running
    onboard the spacecraft.  Under some circumstances, however, a GRB
    may {\it not\/} activate the onboard burst trigger.  For example,
    the burst may be too faint to exceed the onboard detection
    threshold, or it may occur while the onboard burst trigger is
    disabled for technical reasons.  This paper describes a catalog of
    873 ``non-triggered'' GRBs that were detected in a search of the
    archival continuous data from BATSE recorded between 1991 December
    9.0 and 1997 December 17.0.  For each burst, the catalog gives an
    estimated source direction, duration, peak flux, and fluence.
    Similar data are presented for 50 additional bursts of unknown
    origin that were detected in the 25--50 keV range; these events
    may represent the low-energy ``tail'' of the GRB spectral
    distribution.  This catalog increases the number of GRBs detected
    with BATSE by 48\% during the time period covered by the search.
\end{abstract}

\keywords{gamma-rays: bursts}

\section{Introduction}
\setcounter{footnote}{0}
\label{sec:intro}
The Burst and Transient Source Experiment (BATSE) onboard the {\it
  Compton Gamma Ray Observatory\/} is a set of eight uncollimated
detector modules facing outward from the corners of the spacecraft.
Each module has a Large Area Detector (LAD) that is sensitive to
photons in the $\sim$ 25--1000 keV range (\cite{Fishman89}).  The
onboard computer detects gamma-ray bursts (GRBs) and other high-energy
transients in real time.  When the count rates in the LADs exceed a
certain pre-set threshold, the onboard computer signals a ``burst
trigger'' and records data at high temporal and spectral resolution
for a limited time interval (\cite{Fishman89}).  Among the many causes
of onboard burst triggers are GRBs, solar flares, magnetospheric
particle precipitations, bursts from soft gamma-ray repeaters (SGRs),
and variability from X-ray binaries.  The BATSE operations team
continually scrutinizes and catalogs the onboard-triggered events
(\cite{Batse1b,Batse2b,Batse3b,Batse4b,Batse5b}).

Under certain circumstances a GRB or other astronomically interesting
transient may {\it not\/} activate the onboard burst trigger.  The
event may be too faint to exceed the onboard detection threshold, it
may occur while the onboard burst trigger is disabled for technical
reasons, or it may artificially raise the onboard background estimate
and be mistaken for a below-threshold burst.  These events, called
``non-triggered'' or ``untriggered'' bursts, are recorded in the
continuous data that are returned by BATSE.

This paper describes a catalog of 873 non-triggered GRBs that were
detected in a search of the archival data recorded during the 6 years
of the mission between 1991 December 9.0 and 1997 December 17.0.  The
catalog data are also available electronically on the World Wide Web
at {\tt http://space.mit.edu/BATSE/}~.

\section{Onboard Burst Detection}
The real-time burst detection capability on BATSE permits
efficient use of the available telemetry. High-resolution GRB data
are transmitted only in response to a burst trigger.  The
requirements for an onboard burst trigger are programmable.  In
the usual configuration, the count rates from the 50--300 keV
discriminator channels are required to increase above a running
background average by an amount that is specified in terms of the
standard deviation of the expected background counts ($\sigma$) in
each detector.  The required count rate increase is nominally set at
$5.5\sigma$, and at least two detectors are required to
simultaneously exceed this threshold.  The running background
average for each detector is re-computed every 17.408 s.  The
count rates are tested independently on 3 integration intervals:
64 ms, 256 ms, and 1024 ms.

When the onboard trigger criteria are satisfied, the onboard
computer stores data at higher temporal and spectral resolution
for a limited time interval, called the ``accumulation interval.''
The accumulation interval lasts from 180 s to 573 s, during which
further triggering is disabled (\cite{Batse3b}).  Following the
accumulation interval, there is a ``read-out interval.''  During
the read-out interval, data from the burst memories are telemetered
and triggering on the 256 ms and 1024 ms timescales is disabled;
the thresholds for triggering on the 64 ms timescale are set to
the maximum rates of the burst being read out. The duration of the
read-out interval depends on the telemetry schedule, but it
typically lasts from $\sim 28$ minutes for faint burst read-outs
(which omit certain high-time resolution data) to $\sim$ 90
minutes (\cite{Batse4b}).

The monitored energy range, count rate thresholds, and detector
combinations required for an onboard burst trigger are
occasionally re-programmed to pursue certain scientific
investigations or to respond to changes in observing conditions.
This complicates the use of the BATSE GRB catalogs for statistical
analyses that assume uniform detection criteria; but it also
presents new opportunities for analyzing burst properties
(\cite{Batse4b}).

Independently of the onboard burst trigger, the onboard computer
continuously collects lower-resolution data from the LADs.  These
``continuous'' data types are designated in the flight software
and operations work as DISCLA and CONT.  They differ in their time
resolution and in the circuit used to determine gamma-ray photon
energies.  The DISCLA (DISCriminator Large Area) data provide
1.024 s integrations of the count rates in 4 discriminator
energy channels covering the ranges (approximately) 25--50 keV,
50--100 keV, 100--300 keV, and $> 300$ keV.  The CONT (CONTinuous)
data provide 2.048 s integrations of the count rates in 16
pulse-height analyzer channels.  There is a third data type,
DISCLB, that was added in response to the failure of the onboard
tape recorders.  It is just like the DISCLA type except that it
omits the highest discriminator energy channel (\cite{Batse3b}).
The combination of the DISCLA and DISCLB data provide the most
continuous time coverage, although there are gaps owing to missing
or corrupted telemetry and spacecraft passages through the South
Atlantic Anomaly (SAA).

\section{Non-triggered Bursts}
\label{subsec:nontrig} A GRB or other transient may fail to
activate the onboard burst trigger and nevertheless leave a
statistically significant signal in the continuous data.  This can
occur in three circumstances.

First, the burst may not meet the criteria required for the
onboard trigger.  There are three common ways this can happen. The
burst may simply be too faint to exceed the onboard detection
threshold: that is, it may not be bright enough to cause a
5.5$\sigma$ increase in the count rates in at least two detectors.
Another possibility is that the burst occurs during the read-out
of a previous burst, and its peak count rate on the 64 ms time
scale is lower than that of the burst being read out.  Finally,
the burst may have a lightcurve that rises slowly compared to the
17.408 s background averaging. In this case, the burst
artificially raises the nominal background level and thus raises
the detection threshold.  The burst then does not cause an onboard
trigger even though it would have, if a more accurate background
estimate were used.  Such bursts are called ``slow risers.''  The
slowly-rising effect discussed by Lingenfelter \& Higdon (1996)
appears to cause the onboard trigger to miss only a few
percent of GRBs (\cite{Kommers97}).

Second, the burst may occur while the onboard burst trigger is
disabled for technical reasons.  Such bursts may be of any
intensity. The onboard burst trigger is disabled during the
accumulation interval of a previous trigger, so another burst
occurring during this time cannot cause an onboard trigger.  The
onboard trigger is also disabled while the spacecraft passes over
certain geographic locations associated with a high risk of
``false'' triggers caused by precipitation of quasi-trapped
radiation belt electrons (\cite{Batse1b,Datlowe95}).

Third, certain transients will not be detected onboard if they
occur while the trigger criteria are configured for other kinds of
events. For example, a burst or flare from an X-ray binary that
has most of its counts below 50 keV may not be detected onboard
when the trigger criteria are in the nominal 50--300 keV
configuration that is optimized for detecting GRBs.  There are
times during the mission when the onboard trigger criteria were
changed from their nominal settings to study phenomena other than
GRBs, using various energy ranges and detector combinations
(\cite{Batse4b}). During these times sensitivity to GRBs is
impaired.

The archival continuous data contain information on GRBs and other
transients that did {\it not\/} activate the onboard burst
trigger.  They also contain data for many bursts that {\it did}. A
retrospective search of these data can provide a catalog of GRBs
that are found with uniform detection criteria.  It can also can
detect bursts that are fainter than could be studied previously.

\section{The Search for Non-triggered GRBs}
\label{sec:search} This paper is an extension of the search for
non-triggered GRBs described in Kommers et al.\ (1997, hereafter
Paper I).  The goals of Paper I were to evaluate the possibility
of finding non-triggered GRBs and to estimate the rate of
``slowly-rising'' bursts.  The purpose of the present work is to
provide a catalog of non-triggered GRBs that will act as a
supplement to the regular BATSE catalogs.  There are numerous
minor differences between the search and analysis techniques used
for the present work and those used for Paper I; therefore, this
section describes the search techniques in some detail even though
the discussion in Paper I is similar.  The data discussed in Paper
I have been re-analyzed for this catalog.

The search of archival continuous data that is described here is
called the ``off-line search.''  All events detected with the off-line
search are called ``off-line triggers.''  The GRBs detected with both
the off-line search and the onboard burst trigger are called ``onboard
triggered GRBs.''  The GRBs detected {\it exclusively\/} with the
search of archival data are called ``non-triggered GRBs''.

\subsection{Data Selection and Scope}
\label{subsec:data} This catalog covers archival data from 6 years
of the mission, from 1991 December 9.0 to 1997 December 17.0.  The
corresponding Truncated Julian Day (TJD = JD - 2440000.5) numbers
are 8600.0 to 10800.0.  The search makes use of the archival
DISCLA files that are available for 2178 of these 2200 days.
Table~\ref{tbl:missdays} lists the TJD numbers of the days that
are not included in the off-line search.  The total live-time of
the search using the algorithm described in
Section~\ref{subsec:algorithm} is $1.33 \times 10^{8}$ s, which is
70.0\% of the total $1.90 \times 10^{8}$ s spanned by the search.

\subsection{Search Algorithm}
\label{subsec:algorithm}
The basic strategy for the off-line search is much like that of the
onboard trigger: search the data sequentially to identify
statistically significant count rate increases on a given timescale.
There are actually 7 separate off-line searches, each one using a
different combination of discriminator channels and time resolution.
Each search, in turn, uses 2 methods of background estimation.

The first step in each of the off-line searches is to combine the
appropriate energy channels of the DISCLA data to form time series for
each of the eight detectors.  The next step is to search the eight
time series sequentially with a moving window.  For each time bin $k$
where the count rates are available, two different procedures each
yield an estimate of the background rate, $B_d(k)$, in detector $d$.

The first background estimate uses the mean count rates observed
during the 17.408 s preceding the time bin being tested. This estimate
is much like that used onboard the spacecraft, except that it is
continuously recomputed rather than being calculated in discrete
17.408 s blocks. An advantage of this method is that it can be
performed on all the data except for those which follow within 17.408
s of a data gap.

The second method of background estimation takes advantage of the
availability of data recorded after the time bin being tested.  For
the $k$th time bin of the time series the expected number of
background counts in detector $d$ is obtained by interpolating a
linear fit to data in time bins $k - N_b, \ldots , k-1$ and $k + 1 +
N_b, \ldots , k + 2N_b$.  The number of background bins, $N_b$, depends
on the timescale of the search; if the data in time bins $k - N_b$
through $k + 2N_b$ are not temporally contiguous (e.g., if there is a
data gap) then the background estimate is not formed.  Note that the
time bin being tested, $k$, is not centered in the gap of $N_b$ bins
that separates the fitted time intervals.  This somewhat arbitrary
choice is a result of trying various background schemes on 3 weeks
worth of sample data; it gave satisfactory results.  The use of a
longer interval of data for the background fits means that this
background estimate can be formed for fewer time bins than the 17.408
s background estimate.

Assuming the expected background rate $B_d(k)$ is obtained using at
least one of the background estimates, the ``significances'' of the
fluctuations in the detectors on the 1.024 s timescale are defined
from the measured count rates, $C_d(k)$, using
\begin{equation}
S_d(k) = \frac{C_d(k) - B_d(k)}{\sqrt{B_d(k)}}.
\end{equation}
The definition of the ``significances'' for the longer timescales is a
straightforward generalization:
\begin{equation}
S_d(k) = \frac{\sum_{i=0}^{n-1}[C_d(k+i) - B_d(k+i)]}{[\sum_{i=0}^{n-1} B_d(k+i)]^{1/2}}
\end{equation}
Here $n$ is the number of time bins in the timescale of the search.
In the absence of signal and assuming perfect background estimation,
this definition of ``significance'' is intended to yield eight
normally distributed random variables with zero mean and unit variance
associated with time bin $k$.  For $n > 1$ (timescales longer than
1.024 s) the ``significances'' are not independent between time bins
that are separated by fewer than $n$ seconds.

The off-line trigger criteria require that at least two of the
$S_d(k)$ exceed a minimum value of 2.5 and that the sum of the two
greatest exceeds a minimum value of 8.0.  That is, if
$s_1$ and $s_2$ are the greatest and second greatest of the $S_d(k)$,
then the off-line trigger is satisfied when $s_1 \ge s_2 \ge 2.5$ and
$s_1 + s_2 \ge 8.0$.

The parameters of the off-line searches for the 7 combinations time
resolution and energy range are listed in Table~\ref{tbl:srchpar}.
Search A uses counts from the 25--50 keV range at 1.024 s time
resolution, and it is expected to provide optimum sensitivity to short,
low-energy transients such as bursts from SGRs (which typically last
for $\sim 1$ s or less).  Search B uses counts from the nominal GRB
energy channels (50--300 keV) with 1.024 s time resolution for
sensitivity to short GRBs.  Search C uses counts from all three lower
discriminator channels (25--300 keV) to search for signals with a
broad spectral range, such as faint GRBs with substantial low-energy
emission.

The searches on the longer timescales (D through G) provide
sensitivity to longer transients, including the ``slowly-rising''
ones. The GRB duration distribution peaks between 20 s and 50 s so
the statistical significance of most GRBs is expected to be higher
on the longer timescale searches than on the 1.024 s searches.
Thus the longer timescale searches have lower peak flux
thresholds. This is because the significance (as defined above) of
a square-pulse signal in an $n$-second time bin is proportional to
$\sqrt{n}$, assuming the pulse lasts at least $n$ seconds.  The
searches on the 4.096 s and 8.192 s timescales are thus expected
to reach peak fluxes (averaged over the corresponding timescale)
that are lower by factors of (roughly) 2 and 2$\sqrt{2}$,
respectively, than can be reached on the 1.024 s timescale.

The specific time profile of each burst determines which of the three
timescales in Table~\ref{tbl:srchpar} is the most sensitive.  Bursts
detected on one timescale may or may not be detected on another.  For
this reason, each of these searches should be considered a separate
experiment.

The rate of off-line triggers expected from statistical fluctuations
(due to counting statistics alone) is negligible compared to the
actual rate of off-line triggers found in the data.  The probability
that a set of eight significances drawn from the normal distribution
with unit variance will satisfy the off-line trigger criteria can be
estimated by first integrating the bivariate normal distribution
(centered on the origin) over the allowed area $A$ in the ($s_1$,
$s_2$) plane where the inequalities $s_1 \ge s_2 \ge 2.5$ and $s_1 +
s_2 \ge 8.0$ are satisfied (see Figure~1 of Kommers et al.\ 1997).
The resulting probability is then multiplied by the number of ways to
select two detectors out of eight:
\begin{equation}
P_{stat} = \left[ \int\!\!\int_{A} \frac{1}{2\pi} \exp \left( - \frac{s_1^2 +
      s_2^2}{2} \right) \, ds_1 \, ds_2 \right] \times \frac{8!}{6!\ 2!}.
\end{equation}
The value of $P_{stat}$ is $2.1 \times 10^{-7}$ for the 1.024 s
search.  The accuracy of this calculation is verified by Monte Carlo
simulations, which give a probability of $(2.0 \pm 0.5) \times
10^{-7}$.  Thus the rate of ``false triggers'' on the 1.024 s search
due to statistical fluctuations alone is about 0.012 per day of data
searched (assuming 70\% live-time).  Monte Carlo simulations also show
that the 4.096 and 8.192 s searches (in which adjacent time bins are
not independent) give ``false triggers'' at about the same rate.

In practice the off-line trigger detects many more events than can be
expected based on statistical fluctuations.  This is because the time
series being searched are not Poissonian.  They are dominated by the
activity of real sources.  Variable astronomical sources, the Sun,
terrestrial photon sources, and the interaction of cosmic-ray
particles in the detectors all contribute to the count rates.

\subsection{Detection Efficiency and Sky Exposure}
\label{subsec:trigeff}
The ``trigger efficiency'' is the probability that a
transient with given physical characteristics (time profile, photon
flux, energy spectrum, source direction) occurring under given
instrumental conditions (background rates, spacecraft orientation)
will be detected by the off-line trigger.  In the BATSE catalogs, the
trigger efficiency is expressed as function of intensity only, by
averaging over the other relevant quantities.  Similarly, the sky
exposure is the detection probability as function of source direction
only.

The first step in computing the trigger efficiency is to estimate the
probability $E_1(P,\nu,\alpha,\delta,t)$ that an event that occupies
exactly one time bin will be detected.  The event is parameterized by
its peak flux on the timescale of the search ($P$), its power-law
photon spectral index ($\nu$), its source direction (right ascension,
declination) in celestial coordinates ($\alpha$, $\delta$), and time
of occurrence during the mission ($t$).  The time during the mission
determines the background rates and spacecraft orientation, both of
which are obtained from the data files being searched.  If no data are
available at the specified time or if the source direction is occulted
by the Earth, the trigger efficiency is trivially zero.  Otherwise,
the expected count rates in the detectors are obtained using the
instrument and atmospheric response matrices
(\cite{Pendleton95,Pendleton99}).  The expected count rates are
multiplied by the duration of the time bins to obtain the expected
number of counts above background in the detectors, denoted here by
$C^{*}_d$.  The expected background counts $B^{*}_d(k)$ are obtained
from the measured count rates at time $t$ during the mission.  The
expected significances in the detectors are $S^{*}_d = C^{*}_d /
\sqrt{B^{*}_d}$.

Let $s^{*}_1$ and $s^{*}_2$ denote the greatest and second greatest of
the $S^{*}_d$, respectively.  The trigger efficiency
$E_1(P,\nu,\alpha,\delta,t)$ is estimated by integrating the bivariate
normal distribution centered on the point ($s^{*}_1, s^{*}_2$) over
the area $A$ in the $(s_1, s_2)$ plane in which the off-line trigger
criteria are satisfied:
\begin{eqnarray}
\lefteqn{E_1(P,\nu,\alpha,\delta,t) = } \nonumber \\
 & \int\!\!\int_{A} \frac{1}{2 \pi} \exp \left[ -\frac{(s_1 -
    s^{\ast}_1)^{2}}{2} \right] \exp \left[ -\frac{(s_2 -
    s^{\ast}_2)^{2}}{2} \right] \,ds_1 \,ds_2.
\end{eqnarray}
This equation must be evaluated on a grid of peak fluxes, source
directions, spectral indices, and times during the mission.  It is
computationally infeasible to completely sample the ranges of possible
burst characteristics for every time bin checked in the off-line
search.  To obtain a reasonable answer, the efficiency is evaluated on
a grid of 9 peak fluxes, 252 source directions, and a set of times
that sample background rates every 8 minutes on 110 quasi-randomly
chosen days of the mission.  The burst photon spectrum is fixed at a
single typical value, $\nu$ = 2.0.  The peak fluxes span a range where
the efficiency is expected to vary significantly between 0 and 1.  The
source directions are nearly isotropically distributed on the
celestial sphere (\cite{Tegmark96}).

The choice of times at which to compute $E_1(P,\nu,\alpha,\delta,t)$
balances two opposing considerations.  For a given burst, the
efficiency as a function of intensity depends on the background rates.
The variations in background are dominated by orbital modulation, so
there is little need to sample densely in time to get adequate phase
coverage for the distribution of background rates.  On the other hand,
to adequately estimate anisotropies in the sky exposure requires
sampling on a timescale finer than the durations of typical data gaps
due telemetry drop-outs and passages through the SAA.  The compromise
used here is the following.  The first 110 numbers of a Sobol
quasi-random sequence (\cite{Press95}) provide 110 mission days that
uniformly sample the full 2200 days of the off-line search.  On each
of the 110 days, $E_1(P,\nu,\alpha,\delta,t)$ is sampled at time
intervals of 8 minutes.  This scheme should adequately sample the
spacecraft orbital precession period ($\approx 51$ days) and the
orbital modulation of background rates ($\approx 92$ minutes), while
providing a dense sampling to evaluate the effects on sky exposure of
regular data gaps due to the SAA and telemetry schedule.

To obtain $E_1(P)$, the single-time-bin off-line trigger efficiency as
a function of peak flux only, the full trigger efficiency function
$E_1(P,\nu,\alpha,\delta,t)$ is averaged over the quantities $\alpha$,
$\delta$, and $t$ ($\nu$ has the single value 2.0).  This efficiency
underestimates the probability of detecting most GRBs, however,
because most bursts last longer than a single time bin and so have
more than one statistical chance to be detected.

Suppose the peak of a burst occupies $N$ time bins, so that the burst
has effectively $N$ statistical chances to be detected. Then the
probability that the burst is detected can be approximated as unity
minus the probability that the burst {\it fails\/} to be detected in
all $N$ trials:
\begin{equation}
E_N(P) = 1 - \left[ 1 - E_1(P) \right]^{N}.
\end{equation}
Since the number of chances $N$ is not known for GRBs {\it a priori},
the actual probability of detection $E(P)$ is obtained by
marginalizing $E_N(P)$ over the distribution of $N$ for bursts with
peak fluxes near $P$:
\begin{equation}
E(P) = \frac{\sum h(N,P)\,E_N(P)}{\sum h(N,P)}.
\end{equation}
An estimate for $h(N,P)$, the histogram of the various integer values
of $N$ for bursts with peak fluxes near $P$, can be obtained from the
detected sample of bursts by counting, for each burst, the number of
time bins with count rates that were within one standard deviation of
the peak count rate (see \cite{KommersPF}).  The resulting function
$E(P)$ is well represented (to within the uncertainties of the
calculation) by the formula
\begin{equation}
\label{eqn:nteff}
E(P) = \frac{1}{2} \left[ 1 + {\rm erf}(-4.801 + 29.868 P) \right].
\end{equation}
This equation expresses an estimate of the trigger efficiency of our
off-line search on the 1.024 s timescale.  It is plotted as the solid
line in Figure~\ref{fig:nteff}.  The efficiency of the off-line search
falls below 0.5 at a peak flux of 0.16 ph cm$^{-2}$ s$^{-1}$.

The sky exposure of the off-line search is given by averaging
$E_1(P,\nu,\alpha,\delta,t)$ over time to obtain $E_1(\alpha,\delta)$.
The value of $P$ is taken to be well above the detection threshold.
Figure~\ref{fig:skyexp} shows $E_1(\alpha,\delta)$ plotted in
equatorial coordinates.  The maximum exposure is 70.0\%
owing to the presence of data gaps.  The minimum exposure is found
around the celestial equator, $\delta = 0$, owing to blockage by the
Earth.  The Southern Hemisphere has an overall lower exposure than the
Northern Hemisphere owing to data gaps during passages through the
SAA.  Over long time periods, the dependence on right ascension
averages out, so that the sky exposure is effectively a function of
declination only.  The dipole and quadrupole moments of the sky
exposure (in equatorial and Galactic coordinates) are listed in
Table~\ref{tbl:skymom}.

\subsection{Burst Recognition}
\label{subsec:recog} The off-line search procedure discussed in
Section~\ref{subsec:algorithm} detects variability from a variety
of sources.  Most of the off-line triggers are in fact {\it not\/}
associated with GRBs.  They are caused by other impulsive
transients (such as solar flares and bursts from X-ray binaries),
magnetospheric phenomena, and by aperiodic variability from bright
X-ray binaries.  Paper I describes the characteristics that can be
used to classify the various off-line triggers (see also
\cite{Batse1b}).

Briefly, a visual inspection of the count rates in the eight detectors
provides an easy way to identify off-line triggers that are not caused
by astronomical sources of gamma rays.  For example, magnetospheric
particle precipitations create comparable count rates in
opposite-facing detectors or produce inconsistent time profiles
between detectors.  Similarly, phosphorescence spikes due to
excitation of the detector crystal by energetic cosmic ray particles
appear in only one detector.  Egress or ingress of bright X-ray
sources occulted by the Earth and large fluctuations in the aperiodic
variability of X-ray binaries have time profiles that do not resemble
bursts or flares, so these are also easily identified with a plot of
the count rates in each detector.

Burst-like events that appear to be consistent with a point source of
photons can be further examined in the 4 discriminator energy channels
and their source directions can be determined.  This step identifies
solar flares and auroral X-rays based on source direction and
(generally) soft energy spectra.  Bursting and flaring activity from
known X-ray binaries such as GRO~J1744$-$28 can also be
identified based on soft spectra, source direction, and arrival time
during known outbursts of these sources.

The identification of an event as a GRB requires that it show
significant counts in discriminator channels 2 and 3 (50--100 keV and
100--300 keV) and that it show none of the characteristics that might
associate it with the non-GRB causes discussed previously.  Events
that do not show significant signal in the 100--300 keV channel are
not considered GRBs; the reason is that their spectra are softer than
most GRBs and they could therefore be another type of phenomenon, such
as isolated bursts and flares from X-ray binaries.  Such events are
instead placed in an ``unknown'' event category; the catalog of these
``unknown'' events is also given in this paper.

The classification of the off-line triggers by visual inspection is
subjective and occasionally prone to human error.  Fortunately, the
causes of the vast majority of off-line triggers are quite obvious
from the count rate plots and ambiguous cases are rare.  The accuracy
of the burst recognition discussed in Paper I suggests that the
procedures used here will correctly identify $\sim$ 90\% of the
genuine GRBs that are flagged by the off-line trigger and that
contamination of the GRB sample by non-GRB events will occur at the
level of $\sim$ 2\% of the size of the combined triggered plus
non-triggered GRB sample.  The accuracy of the off-line trigger
classification is discussed further in Section~\ref{subsec:class}.

\subsection{Off-line Triggers}
\label{subsec:offline}
The total number of off-line triggers detected in the search of 6
years of archival data is 371,474.  This large number does not
directly reflect the number of individual transients that are
responsible for the off-line triggers, however.  A single transient
can cause from one to hundreds of off-line triggers.  For example, a
typical $\sim 10$ s duration GRB may cause 2 to 20 off-line triggers,
while a magnetospheric electron precipitation lasting several minutes
may cause several hundreds of off-line triggers.

The rate of off-line triggers varies dramatically according to the
activity of the Sun and certain bright X-ray binaries.
Figure~\ref{fig:alldets} shows ten-day averages of the rate of
off-line triggers.  The prominent features have been labeled according
to the sources responsible for most of the detections.  The time
period used to experiment with different off-line trigger algorithms
and thresholds (TJD 9200--9221) is unfortunately one of the
``quietest'' time periods in the data.  Higher thresholds might have
been chosen if this had been known in advance.

The vast majority of off-line triggers are not associated with
GRBs. Most of the off-line triggers are associated with aperiodic
variability from bright X-ray binaries in outburst and atmospheric
particle precipitations.  Table~\ref{tbl:breakdown} shows a
breakdown of the causes of off-line triggers as determined from
the visual inspection described in section \ref{subsec:recog}.
This table shows that only about 17\% of the off-line triggers are
associated with burst-like astronomical transients: GRBs, solar
flares, bursts from the X-ray binary GRO~J1744$-$28, bursts from
SGRs, and transients of unknown origin.

\section{The Non-triggered GRB Catalog}
\label{sec:catalog}
There are 2265 bursts (corresponding to $\sim$ 22,000 off-line
triggers) that were classified by visual inspection as GRB candidates.
Of these, 1392 are onboard triggers that are already listed as GRBs in
the BATSE trigger catalog (\cite{Batse5b}).  The onboard trigger
numbers for these events are listed in Table~\ref{tbl:triggrb}.  The
remaining 873 events that did not activate the onboard burst trigger
are the non-triggered GRB candidates.  The times, source directions,
durations, peak fluxes, and fluences for these events are listed in
Table~\ref{tbl:ntgrb}.  The union of the 1392 onboard triggered
bursts with the 873 non-triggered bursts constitutes a well-defined
sample of GRBs found with homogeneous detection criteria.

During the time period covered by the off-line search (TJD 8600.0
through TJD 10800.0) the onboard burst trigger detected 1815 events
that are listed in the BATSE trigger catalog as GRBs.  The off-line
search did {\it not\/} detect 423 of these onboard triggers owing to
gaps in the DISCLA data and to the fact that the 1.024 s time resolution
of the data reduces the statistical significance of bursts with
durations shorter than about 1 s.

\subsection{Estimation of Physical Parameters}
\label{subsec:phyparam}
The non-triggered GRB catalog (Table~\ref{tbl:ntgrb}) is a supplement
to the regular BATSE burst catalogs.  The physical parameters listed
in it are therefore those that the BATSE team reports for
onboard-triggered GRBs (\cite{Batse1b,Batse3b,Batse4b,Batse5b}).
These quantities are source direction, duration, peak photon flux,
fluence, and the ratio of the maximum count rate observed during the
burst to the minimum count rate required for detection.

The first step in estimating the physical parameters of a burst is to
determine background count rates.  Using interactive plotting and
analysis software, background time intervals are selected from before
and after the burst.  The data in the selected intervals are used to
determine a linear least-squares best-fit polynomial.  The background
during the burst is assumed to be the interpolation of this polynomial
to the times between the selected intervals.  The degree of the
polynomial depends on how the background appears to be changing on the
timescale of the burst.  Typically a polynomial of order 4 or less
provides both a good fit to the selected background data and shows
reasonable behavior in the intermediate interval where the burst
occurs.  This procedure is the same as the one used by the BATSE
analysis team.

\subsubsection{Source Directions}\label{locburst}
To estimate the source direction, a time interval containing the
``main'' part of the burst is selected by hand.  The purpose of
this selection is to obtain optimum signal-to-noise ratios for the
average count rates during the burst.  The mean
background-subtracted count rates obtained from the main section
of the burst are used to derive a best-fit source direction.  The
software for this purpose uses the detector and atmospheric
response matrices to predict the count rates expected from an
event with given physical parameters: the intensity $P$ measured
in ph cm$^{-2}$ s$^{-1}$ in a given energy range, the power-law
photon spectral index $\nu$, and the source direction in terms of
the azimuthal and elevation angles $(\phi, \theta)$ in {\it
CGRO\/} coordinates.  The best-fit values for these quantities are
determined by minimizing the following measure of goodness-of-fit:
\begin{equation}
\chi^2 = \sum_{d} \sum_{i} \left[ \frac{C^{i}_{d} -
    \tilde{C}^{i}_{d}(P,\nu,\phi,\theta)}{\sigma_{d}^{i}} \right]^2,
\label{eqn:chi2min}
\end{equation}
where $C^{i}_{d}$ denotes the measured count rates in detector $d$ and
discriminator channel $i$, $\sigma_{d}^{i}$ denotes the uncertainties
in the measured count rates, and
$\tilde{C}^{i}_{d}(P,\nu,\phi,\theta)$ denotes the count rates
predicted by the response matrices.  The sum over detectors is
restricted to the 4 detectors that face the given source direction.
The sum over energy channels is generally restricted to DISCLA
channels 2 and 3 (50--300 keV) for GRB candidates and to channels 1
and 2 (25--100 keV) for bursts from SGRs and other low-energy transients.

The geometry of the BATSE detectors can allow for multiple, widely
separated local minima in the $\chi^2$ parameter space.  This is
especially true for weak bursts that have little or no signal in the
third and fourth most brightly illuminated detectors.  When most of
the counts are in just two detectors, the data can be equally
consistent with between 2 and 4 source directions depending on the
choice of the third and fourth most brightly illuminated detectors.
In the absence of strong signal this choice can be strongly influenced
by fluctuations in the count rates due to counting statistics or
variability from unrelated background sources.  For this reason, the
uncertainties in the source directions are determined using a
boot-strap Monte Carlo approach.  A set of 50 synthetic count rates is
generated by sampling from Poisson distributions with the same means
and variances as the measured count rates.  The best fit locations are
determined for these synthetic count rates.  The uncertainty in the
source direction is the radius of the circle on the unit sphere that
encloses 34 (68\%) of the synthetic best-fit source directions.

\subsubsection{Durations}
The durations of events are measured using the $T_{50}$ and $T_{90}$
statistics (\cite{Kouveliotou93}).  These duration measures are the time
intervals during which the total number of counts in the burst
increases from 25\% to 75\% and from 5\% to 95\%, respectively, of the
total counts.  The algorithm for computing them and an evaluation of
the systematic errors arising from the background modeling are given in
Koshut et al.\ (1996).  The intervals chosen for the
background fit and the order of the polynomial used to interpolate the
background during the burst are subjective, so systematic errors arise
depending on the burst profile.  A particularly problematic case arises
for bursts that have extended, low level emission that is difficult to
distinguish from background variations.  The magnitudes of the
systematic errors in the $T_{90}$ duration measure can then be as
large as $\sim 20$\% (\cite{Koshut96}).

For some events, particularly faint ones, background fluctuations
cause the durations estimated by the $T_{50}$/$T_{90}$ algorithm to be
obviously unreliable.  These cases are easily identified by visual
inspection of the count rate plots that have the boundaries of the
duration measures superimposed.  In cases where the algorithm fails,
an estimate of $T_{90}$ and its associated uncertainty is made by
visual inspection of the burst profile.

\subsubsection{Peak Photon Fluxes}
The photon spectrum of an event is assumed to be a simple
power-law over the energy range of interest (50--300 keV for GRB
candidates, 25--100 keV for low-energy events).  The best-fit peak
photon flux is determined by minimizing equation \ref{eqn:chi2min}
with respect to $P$ and $\nu$.  The count rates used for the
minimization are the count rates observed in the detectors during
the 1.024 s time bin for which the count rate (above background)
in the two most brightly illuminated detectors is a maximum.  The
source direction is fixed at the best-fit value found previously
using the (often) more precise count rates from the ``main''
portion of the burst.

This procedure is somewhat different from that used to derive the
peak fluxes listed in the BATSE catalogs.  For bursts that caused
an onboard trigger, data types are available with time resolution
better than 1.024 s.  The peak fluxes in the BATSE catalogs make
use of data with 64 ms time resolution to find the optimum
placement of a 1.024 s window that yields the highest count rate
averaged over that window (\cite{Pendleton96}).  In contrast, the
DISCLA data used for the off-line analysis do not allow the
freedom to choose the phase of the 1.024 s window that contains
the most counts from the burst.  In the worst case, the 1.024 s
window used by the BATSE team may get split into two DISCLA time
bins, so the peak flux determined with the off-line search will be
lower by a factor of $\sim 2$.  The BATSE catalogs also use a more
sophisticated model of the GRB photon spectrum when estimating
peak fluxes and fluences (\cite{Pendleton96}).

\subsubsection{Fluences}
The estimates of burst fluence make use of the count rates (above
background) from the $T_{90}$ interval.  The fluence $F$ is
estimated by the formula $F = cTP\langle E \rangle$, where $c =
1.602 \times 10^{-9}$ erg keV$^{-1}$ is a conversion factor from
keV to erg, $T$ is the duration of the time interval over which
the fluence is estimated (usually $T_{90}/0.9$), $P$ is the
average photon flux over the time interval $T$, and $\langle E
\rangle$ is the mean energy of the photon spectrum in the
energy range of interest. GRBs often show some degree of spectral
softening over their duration, so events with $T_{90}$ longer than
4.096 s are partitioned into 4.096 s sub-intervals and the total
fluence is obtained by summing the fluences estimated for each
sub-interval.  For bursts with $T_{90}$ less than 4.096 s, the
$T_{90}$ interval itself is used.

\subsubsection{$C_{\rm max}/C_{\rm min}$}
To compute the ratio of the maximum count rate detected during the
burst to the minimum count rate required for detection is more
complicated for the off-line trigger than for the onboard trigger.
The conditions for an onboard trigger depend on the count rates in a
single detector: the one with the second highest number of counts
above the background estimate.  In contrast, the off-line trigger
depends not only on the values of the count rates in two detectors but
on their ratio.  Let $s_1$ and $s_2$ denote the significances of the
most significant and second most significant detector, respectively.
The source intensity required for detection could be determined either
by the point at which $s_2$ falls below 2.5 or the point at which the
sum $s_1 + s_2$ falls below 8.0; which of these alternatives obtains
is determined by the ratio of $s_1$ to $s_2$.

\subsection{Accuracy of Off-line Burst Parameters}
A sample of 80 onboard-triggered GRBs that were also detected with the
off-line search provides a basis for evaluating the accuracy of the
burst parameters derived using the techniques described previously.
The source directions, peak fluxes, fluences, and durations derived
using the off-line search techniques can be compared to the values
derived by the BATSE team (\cite{Batse5b}).

A measure of the disagreement between the estimates produced by
each method is provided by the difference in best-fit values
divided by the uncertainty in the difference arising from the
combined statistical and systematic error estimates
associated with each measurement.  The formula for the
``measurement disagreement'' $\Delta$ is:
\begin{equation}
\Delta = \frac{X_B - X_O}{\sqrt{\sigma_B^2 + \sigma_O^2 + \sigma^2_{\rm sys}}},
\end{equation}
where $X_B$ and $X_O$ denote the values of a physical quantity
estimated by the BATSE team and by the off-line search methods,
$\sigma_B$ and $\sigma_O$ denote the estimates of the statistical
uncertainties in each estimate, and $\sigma_{\rm sys}$ denotes
estimated contributions to the uncertainties from systematic
errors.  In the case of source directions, the arclength on the unit
sphere is used as the ``difference'' between the source directions.
Figure~\ref{fig:calib} shows histograms of the measurement
disagreement for the 80 onboard-triggered bursts used in the
comparison.  Dotted histograms do not include any contribution from
random systematic errors ($\sigma_{\rm sys} = 0$), while solid
histograms do.  The figure shows that the physical quantities
estimated with the off-line search methods agree reasonably well with
those estimated by the BATSE team.

The direction estimates in the BATSE catalog are estimated to contain
a systematic uncertainty of 1.6$^{\circ}$ to be added in quadrature to
the quoted statistical uncertainties; that is, the uncertainty in the
direction to a burst with a 2.0$^{\circ}$ statistical error is
$\sqrt{1.6^2 + 2.0^2}$ degrees (\cite{Batse4b}).  The magnitude of the
systematic errors is estimated by comparison with transients from
sources with well-known coordinates, such as Cyg X-1, the Sun, and
GRBs localized by arrival-time triangulation with the Interplanetary
Network.  Using a suite of solar flares and X-ray bursts from
GRO~J1744$-$28 the systematic uncertainty in the off-line search
source directions is estimated to be approximately 4$^{\circ}$, to be
added in quadrature to the statistical uncertainties reported in the
catalog.  The systematic error contributions used to make the solid
histogram for the source direction comparison in
Figure~\ref{fig:calib} contain both the 1.6$^{\circ}$ and the
4$^{\circ}$ contributions to $\sigma_{\rm sys}$.  With systematic
uncertainties included, 76\% of the measurement disagreements are less
than $\Delta = 1.0$, and 94\% are less than $\Delta = 2.0$.

The peak fluxes listed in the BATSE catalogs have systematic
uncertainties on the order of 20\% (\cite{Batse4b}).  If this 20\%
random systematic uncertainty is added in quadrature to the
statistical uncertainties of the off-line peak fluxes, then 76\% of
the bursts have measurement disagreements less than $\Delta = 1.0$ and
93\% have disagreements less than $\Delta = 2.0$.

The fluences are particularly prone to systematic errors.  They depend
on the durations of the bursts, which may be hard to determine if
there are long ``tails'' of low level emission that could be confused
with background.  They also depend on the model of the photon
spectrum, and the simple power-law that is used in the off-line
analysis may be inadequate in some bursts.  With a 20\% uncertainty
added in quadrature to the statistical uncertainties in the off-line
fluences, 55\% of the bursts have measurement disagreements less than
$\Delta = 1.0$ and 79\% have disagreements less than $\Delta = 2.0$.

The durations are also prone to systematic errors related to the
choice of background intervals.  For many bursts it is difficult to
distinguish low-level emission from background variations.  If a 20\%
systematic uncertainty is added in quadrature to the statistical
uncertainties of the off-line durations, 56\% of the durations have
measurement disagreements less than $\Delta = 1.0$ and 73\% have
disagreements less than $\Delta = 2.0$.

\subsection{Accuracy of Burst Recognition}
\label{subsec:class}
There are two kinds of errors that can occur when deciding if a given
off-line trigger is a GRB or not.  The first is to identify an event
as a GRB when it is in fact caused by some other phenomenon.  The
second is to classify a genuine GRB as something else.

\subsubsection{Contamination by Non-GRB Events}
The non-triggered GRBs can be examined for contamination by solar
flares and magnetospheric particle precipitation events.
Figure~\ref{fig:sunhist} shows the histogram of angles between the
Sun and the source direction for the non-triggered GRBs.  The
solid curve shows the expectation for GRB directions distributed
randomly with respect to the solar direction.  This histogram
shows no evidence for an excess of events in the direction of the
Sun, nor for a substantial deficit (but see
section~\ref{sec:grbdir}).

Significant contamination of the non-triggered GRB sample by auroral
X-rays (magnetospheric particle precipitation events occurring at some
distance from the spacecraft) is expected to appear as an excess of
events detected from directions near the limb of the Earth.
Figure~\ref{fig:geoang} shows the histogram of angles between the
direction of the center of the Earth and the source directions of the
non-triggered GRB candidates.  The dotted line shows the approximate
location of the Earth's limb (the exact angle varies with the altitude
of the spacecraft).  The distribution of angles does not cut off
sharply at the Earth's limb because of the smearing effect of
occasionally large direction uncertainties.  There is no evidence in
this figure for a significant contribution from auroral X-ray events
in the non-triggered GRB sample.

\subsubsection{Incompleteness by Mis-classification of GRBs}\label{subsec:miss}
Even if it is rare that an event gets classified as a GRB when it is
really something else, the converse possibility still exists: that
some fraction of genuine GRBs get classified as some other kind of
phenomenon (such as solar flares or particle precipitations).  In the
absence of an independent set of event classifications, one way to
estimate how many GRBs have been classified as something else is to
review the off-line triggers and their classifications to look for
ambiguous cases or examples of simple human error.

For 77 days of data distributed throughout the mission, a secondary
review of the off-line triggers reveals 3 events that are likely to be
GRBs, but that were classified as something else (owing to human error
or to ambiguity in the characteristics of the event).  The total
live-time of the search during these 77 days is $4.6 \times 10^{6}$ s.
To estimate the number of GRBs missed (due to mis-classification) in
the entire 6 years of the search, it is assumed that the rate of
mis-classified bursts is constant and that bursts are mis-classified
according to the Poisson distribution.  Then the likelihood function
(\cite{LoredoStat}) for the rate of missed bursts when 3 events are
observed in $4.6 \times 10^{6}$ s has a maximum at $6.5 \times
10^{-7}$ s$^{-1}$.  This corresponds to a most likely value of 86
genuine GRBs that have been classified as non-GRB phenomena in the
entire off-line search.  The limits of the central 90\% confidence
interval indicate that between 50 and 200 bursts have been
mis-classified.  This corresponds to a ``loss rate'' for GRBs of
between 2\% and 8\% (most likely 4\%) of the 2265 detected in the
off-line search.

\section{GRB Direction Distribution}\label{sec:grbdir}
The direction distribution of the non-triggered GRB candidates is
shown in Figure~\ref{fig:grbsky}.  There are a number of
statistics to test the hypothesis that these directions are
isotropic (\cite{Briggs96}).  Table~\ref{tbl:isostat} lists the
dipole and quadrupole moments in equatorial and Galactic
coordinates for the non-triggered GRB candidates alone and for the
combined sample of triggered plus non-triggered GRBs that were
detected by the off-line search.  The moments in the table have
been corrected for the sky exposure shown in
Figure~\ref{fig:skyexp}. The dipole and quadrupole moments in
Galactic coordinates are sensitive to concentrations towards the
Galactic center and the Galactic plane, respectively.  The dipole
and quadrupole moments in equatorial coordinates are sensitive to
concentrations towards Earth's North pole and towards Earth's
poles, respectively (\cite{Briggs96}).  There is no significant
evidence for anisotropy in these samples.

The faintest bursts detected with the off-line search are those below
the onboard detection threshold.  A sample of faint GRBs can be
defined by taking only those bursts with peak fluxes (on the 1.024 s
timescale in 50--300 keV) that are less than 0.3 ph cm$^{-2}$
s$^{-1}$, the approximate onboard threshold for 50\% detection
probability (\cite{Batse4b}).  In this sample there are 669 bursts, of
which 551 are strictly non-triggered GRBs and 118 were also detected
with the onboard trigger.  The dipole and quadrupole moments for this
sample are given in Table~\ref{tbl:isostat}.  They show marginal
($2\sigma$) evidence for clustering near the poles of the equatorial
coordinate system.  This is almost certainly due to systematic
classification errors.  There is a tendency to classify events that
arrive from source directions consistent with the Sun as solar flares.
This is because a faint GRB arriving from a direction near the Sun is
more likely to have characteristics such that it would be difficult to
argue that it is {\it not\/} a solar flare.  Thus some GRBs will be
``lost'' from the non-triggered GRB catalog because they are
classified as solar flares.

\section{GRB Duration Distribution}
The $T_{90}$ durations of the non-triggered GRB candidates range
from 1.024 s (the minimum that can be measured with DISCLA data)
to 280 s. The longest burst detected with the off-line search is
one that caused an onboard trigger and has $T_{90} = 674$ s
(\cite{Batse5b}). Figure~\ref{fig:untdur} shows the duration
distributions of the non-triggered GRB candidates only (solid
histogram) and of the onboard-triggered bursts that were detected
with the off-line search (dashed histogram).  The durations of the
non-triggered bursts are necessarily equal to or longer than the
1.024 s time resolution of the DISCLA data.  The onboard-triggered
bursts have durations listed in the current BATSE catalog
(\cite{Batse5b}) that are determined from the higher time
resolution data available in response to an onboard trigger. The
similar shapes of the two curves for durations exceeding $T_{90} =
2.048$ s indicate that the bursts detected with the off-line
search have approximately the same duration distribution as those
detected onboard.  The figure also shows that the off-line search
has difficulty detecting bursts with durations of $T_{90} \approx
1.024$ s or less.  Evidently the off-line search is not detecting
any new burst population based on duration.

\section{GRB Intensity Distributions}
The intensity distributions of the GRBs detected with the off-line
search are discussed in detail in Kommers et al. (2000).  Of the
various measures of burst intensity given in
Table~\ref{tbl:ntgrb}, the peak flux distribution is likely to be
the most straightforward to interpret.  This is because it is
given in physical units and can be obtained with reasonable
accuracy from the raw count rate data.  The maximum count rates
encountered on the 1.024 s, 4.096 s, and 8.192 s timescales are
simple measures of burst intensity, but they are not in physical
units and they depend on the details of the detector response
functions (\cite{Pendleton96}).  The fluence is meant to reflect
the total energy deposited in the detectors by the burst, but
spectral evolution during the burst and uncertainty regarding its
duration make the fluence more difficult to extract from the data.
As shown in Figure~\ref{fig:calib} the peak fluxes are more
reliably estimated from the raw count rate data than are the
fluences.

The value of the $\langle V/V_{\rm max} \rangle$ statistic in
Euclidean space is estimated from the count rate data by computing the
value of $\langle (C_{\rm min}/C_{\rm max})^{3/2} \rangle$, where
$C_{\rm min}$ is the minimum count rate required to detect each burst
and $C_{\rm max}$ is the maximum count rate recorded during the burst
(\cite{Schmidt88}).  For all the bursts detected in the off-line
search, the value of $\langle (C_{min}/C_{max})^{3/2} \rangle$ is
$0.177 \pm 0.006$.  For the bursts detected exclusively on the 1.024 s
timescale, $\langle (C_{min}/C_{max})^{3/2} \rangle = 0.247 \pm
0.006$.

For further information regarding the value of $\langle
(C_{min}/C_{max})^{3/2} \rangle$ and the interpretation of the peak
flux distribution in the context of cosmological models for the burst
sources, see Kommers et al.\ (2000).

\section{``Unknown'' Low-energy Events}
Events that do not have the obvious signatures of terrestrial
phenomena, instrumental effects, solar flares, GRBs, bursts from SGRs,
or bursts from X-ray binaries are put into the ``unknown'' category.
Membership in this category does not rule out any of these causes,
however.  It just means that there is not enough information to make a
confident identification. Thus the unknown category is almost
certainly a mix of events from a variety of sources.  In fact, of the
125 events that originally constituted this class, only 50 remain
listed.  The 75 discarded events appear in retrospect (upon closer
examination) to be due to magnetospheric particle precipitations or to
aperiodic variability from known X-ray binaries in outburst.  This
kind of revision shows that the ``unknown'' low-energy events are
identified less reliably than those classified as GRBs and SGRs.  The
50 ``unknown'' low-energy events are listed in Table~\ref{tbl:unktbl}.

The direction distribution of the 50 low-energy events is
consistent with isotropy.  Figure~\ref{fig:unksky} shows sky maps
of the best-fit source directions in equatorial coordinates. There
is no obvious clustering of burst directions.
Table~\ref{tbl:unkiso} gives the values of the dipole and
quadrupole statistics for these events.

The unknown events are selected to be those bursts that are
spectrally softer than is typical for GRBs but that cannot
obviously be assigned to some other cause, such as a
magnetospheric particle precipitation or solar flare.  A crude way
to characterize the spectral characteristics of these events is
with an instrumental hardness ratio, the ratio of the mean count
rates in two energy bands.  Figure~\ref{fig:unkhard} shows
hardness ratios for the non-triggered GRB candidates (points) and
for the ``unknown'' events (diamonds with error bars).  The
horizontal axis shows the ratio of mean count rates in
discriminator channel 1 to those in the sum of channels 1 and 2.
Likewise, the vertical axis shows the the ratio of mean count
rates in discriminator channel 3 to the sum of those in channels 2
and 3.  The ``unknown'' events are spectrally softer because they
were chosen to be.  The large uncertainties associated with their
position on this diagram follows from the fact that they are
typically too faint to classify with confidence.

The arrival times of the ``unknown'' events are somewhat irregular.  A
K-S test rejects with high confidence the null hypothesis that
``unknown'' events arrive at the detector at a constant rate.
Assuming the null hypothesis that the rate of detected ``unknown''
events is proportional to the live-time of the search (as would be
expected if they arrive with a constant rate), the probability of
getting a higher value of the K-S statistic ($D = 0.28$) is $6 \times
10^{-4}$.  The irregularities in the rate of these events are most
likely due to a combination of variations in observing conditions,
inconsistencies in the classification of events, and irregular
bursting or flaring activity from the sources contributing to this
class.

The unknown class of events probably includes bursts from a variety of
unrelated phenomena.  Too little information is available to
confidently identify them.  One possibility is that many of them
represent the low-energy ``tail'' of the GRB spectral distribution.
The off-line trigger classification process guarantees that the
softest GRBs (those with little or no significant counts above 100
keV) will be placed in the unknown category unless they resemble a
solar flare or particle precipitation event.  The isotropic direction
distribution and the spectral characteristics of the unknown events
are consistent with this hypothesis.  Their arrival times are not
uniform, however, and the rate of spectrally soft GRBs is not expected
to vary over the mission; but again, this may be due to changes in
observing conditions or inconsistencies in the event classification.

\section{Summary}
\label{sec:summary}
A search of 6 years of archival continuous data from BATSE detects
2265 GRB candidates, of which 873 were {\it not\/} detected with the
onboard burst trigger.  This sample of GRBs reaches peak fluxes on the
1.024 s timescale that are a factor of $\sim 2$ lower than could be
studied previously.  The direction distribution of the non-triggered
GRBs is consistent with an isotropic source distribution.  The
durations of long ($T_{90} > 2$ s) bursts are representative of those
in the regular BATSE GRB catalogs.  The 50 bursts of ``unknown''
origin detected in the search of the 25--50 keV data probably
represent a variety of sources; many of them are expected to represent
the low-energy tail of the GRB spectral distribution.

\section{Note Added in Manuscript}
\label{sec:note}
Since this catalog was completed (March 1999) Stern et al. (1999, 2000a, 2000b)
published results of another search of the archival BATSE data for
non-triggered GRBs.  Their search uses off-line trigger criteria that
are nearly identical to those discussed in this paper.  Comparison of
the GRBs listed in a preliminary version of the Stern et al.\ catalog
(B. Stern, 1999, private communication) with the GRBs listed in this
paper revealed that there are at least 213 non-triggered GRBs that are
omitted from Table \ref{tbl:ntgrb} of this catalog owing to
mis-classification of the events as non-GRB transients.  This number
of mis-classifications is comparable to the upper limit of the range
estimated in section \ref{subsec:miss}.  The effect of this
discrepancy on the peak flux distribution is discussed further in
Kommers et al. (2000).

Similarly, Stern et al. (2000b) report that there are at least 90
confirmed GRBs listed in this catalog (Table \ref{tbl:ntgrb}) that are
not included in their sample.  This illustrates the limitations of
burst recognition in both searches, and this uncertainty carries over into
various catalogs of GRBs in the form of sample incompleteness and
contamination by non-GRB transients.  A more complete GRB sample would be
obtained from the union of the events in the two catalogs.

In addition to the uncertainty of burst recognition, Stern et al.\
(2000a) obtain a GRB peak flux distribution that is steeper at the
faint end than the distribution obtained from this catalog.  This
discrepancy is too large to blame simply on the differences in burst
recognintion.  Possible causes may include the differences in data
reduction procedures for translating raw counts into physical peak
flux units.  The procedures described by Stern et al.\ (2000b) make
use of a more highly parameterized spectral form than the simple
power-law used for this work.  A detailed investigation of these
discrepancies is beyond the scope of this paper.

\acknowledgments J. M. K. acknowledges support from NASA Graduate
Student Researchers Program Fellowship NGT8-52816.  W. H. G.  L.
acknowledges support from NASA under grant NAG5-3804. C. K.
acknowledges support from NASA under grants NAG5-32490 and NAG5-4799.
J. v.\ P.  acknowledges support from NASA under grants NAG5-2755 and
NAG5-3674.

\clearpage


\clearpage


\begin{figure}
\caption{Trigger efficiency for our off-line search.  The grid points
  of the calculations are plotted as individual symbols.  Error bars
  represent the standard deviations of the calculated probabilities
  owing to variations in the background rates.  The long-dashed line
  shows the probability that a burst occupying a single time bin is
  detected by our search.  The solid line shows the marginal
  probability that a burst is detected by our search, given that some
  bursts longer than 1.024 s have more than one statistical chance to
  be detected.  For comparison, the dotted line shows the trigger
  efficiency from the 4B catalog; no uncertainties are available for
  the grid points (squares).}
\label{fig:nteff}
\end{figure}

\begin{figure}
\caption{The sky exposure of the off-line
  trigger, expressed as percentage of elapsed time in equatorial
  coordinates.  The maximum exposure is 70.0\% owing to gaps in the
  DISCLA data.}
\label{fig:skyexp}
\end{figure}

\begin{figure}
\caption{The total rate of off-line
  triggers, shown as 10-day averages.  Aperiodic variability from
  X-ray binaries in outburst is responsible for the most prominent
  peaks.  Selected features are labeled according the source
  responsible.}
\label{fig:alldets}
\end{figure}

\begin{figure}
\caption{Histograms of the relative disagreement in measurements
  of burst properties.  The measurement disagreement ($\Delta$) is
  defined as the difference between the off-line estimate and the
  BATSE team's estimate divided by the uncertainty in the difference.
  The uncertainty in the difference combines the statistical
  uncertainties of each measurement in quadrature with an estimate of
  random systematic uncertainties.  Solid histograms include
  systematic uncertainties; dotted histograms do not.}
\label{fig:calib}
\end{figure}

\begin{figure}
\caption{Histogram of angles between non-triggered GRB sources and the
Sun.  The solid curve shows the expectation for direction distributed
randomly with respect to the direction of the Sun, as is expected for
GRBs.  There is no evidence for an excess of
non-triggered GRB candidates in the direction of the Sun; there is,
however, some indication for a deficiency owing to the tendency to
classify some faint GRBs as solar flares when they have directions
consistent with the sun (see section 6).}
\label{fig:sunhist}
\end{figure}

\begin{figure}
\caption{Histogram of angles between non-triggered GRB sources
  and the Earth's center.  The solid line shows the expectation for
  direction distributed randomly with respect to the direction of the
  Earth's center, as is expected for GRBs.  There is no evidence for a
  large excess of directions near the Earth's limb, as would be
  expected if there were substantial contamination from auroral
  X-rays.}
\label{fig:geoang}
\end{figure}

\begin{figure}
\caption{The best-fit  source directions for the non-triggered GRB candidates.  The median
  direction uncertainty is \protect{6$^{\circ}$}.}
\label{fig:grbsky}
\end{figure}

\begin{figure}
\caption{The duration
  distribution of GRBs detected with the off-line search.  The dashed
  line indicates the duration histogram for GRBs that triggered
  onboard.  The solid line indicates the duration distribution of the
  non-triggered GRB candidates.  The shapes are very similar.  This
  shows that the off-line search is not detecting a population of
  bursts with durations significantly different from those detected
  onboard.}
\label{fig:untdur}
\end{figure}

\begin{figure}
\caption{The best-fit
  source directions for the ``unknown'' low-energy bursts.  The median
  direction uncertainty is \protect{16$^{\circ}$}.}
\label{fig:unksky}
\end{figure}

\begin{figure}
\caption{Hardness ratios of ``unknown'' events (diamonds with error bars)
  and non-triggered GRBs (points).  The ``unknown'' events are
  selected to be softer than GRBs, but they are not far separated from
  the population of GRB hardness ratios.  For clarity, error bars are
  not shown for the GRBs.}
\label{fig:unkhard}
\end{figure}

\clearpage

\begin{deluxetable}{ccc}
\tablehead{\multicolumn{3}{c}{Missing Days (TJD)}}
\tablecaption{Days for which no archival data were searched.
\label{tbl:missdays}}
\startdata
  ~8947 &  10322  & 10330 \nl
  ~9111 & 10323   & 10331 \nl
  ~9112 &  10324  & 10332 \nl
  ~9114 & 10325   & 10333 \nl
  ~9153 & 10326   & 10334 \nl
  ~9474 & 10327   & 10335 \nl
  ~9827 & 10328   & 10336 \nl
    ~ &    10329 & ~     \nl
\enddata
\end{deluxetable}

\begin{deluxetable}{cclccrc}
\tablehead{~ & Timescale & Discriminator \\
Search & (s) & Channels}
\tablecaption{Parameters of Off-line Searches
\label{tbl:srchpar}}
\startdata
A & 1.024 & 1 \nl
B & 1.024 & 2 and 3 \nl
C & 1.024 & 1, 2, and 3 \nl
D & 4.096 & 1 \nl
E & 4.096 & 2 and 3 \nl
F & 8.192 & 1 \nl
G & 8.192 & 2 and 3 \nl
\enddata
\end{deluxetable}

\begin{deluxetable}{lllr}
\small
\tablehead{Statistic\tablenotemark{a} &
  \begin{tabular}{c}Coordinate\\System\end{tabular} &
  Moment &
  Value\tablenotemark{b}}
\tablecaption{Dipole and quadrupole moments of sky exposure
\label{tbl:skymom}}
\startdata
$\langle \cos \theta\ \rangle$ & Galactic & Dipole & $-0.0100$\nl
$\langle \sin^2 b - \frac{1}{3} \rangle$ & Galactic &
Quadrupole & $-0.0049$ \nl
$\langle \sin \delta \rangle$ & Equatorial & Dipole & $0.0219$ \nl
$\langle \sin^2 \delta - \frac{1}{3} \rangle$ & Equatorial &
Quadrupole & $0.0240 $ \nl
\enddata
\tablenotetext{a}{$\theta$ is the angle between the source direction
  and the Galactic center, $b$ is the Galactic latitude, and $\delta$ is
  the declination.}
\tablenotetext{b}{Uncertainties in the moments (owing to uncertainties
  in the sky exposure calculation) are estimated to be $\pm 0.0002$.}
\normalsize
\end{deluxetable}

\begin{deluxetable}{lr}
\tablehead{Cause of Off-line Trigger & Percentage of Off-line Triggers}
\tablecaption{Some Causes of Off-line Triggers
\label{tbl:breakdown}}
\startdata
Aperiodic variability from X-ray binaries & ~ \nl
~~~and particle precipitations & 70 \nl
Occultation steps & 13 \nl
GRBs & 6 \nl
Solar flares & 4 \nl
GRO~J1744$-$28 & 6 \nl
Other (including SGRs, low-energy unknown events)& 1 \nl
\enddata
\end{deluxetable}

\begin{table}
  \dummytable\label{tbl:triggrb}
\end{table}

\begin{table}
  \dummytable\label{tbl:ntgrb}
\end{table}

\begin{deluxetable}{llllr}
\small
\tablehead{Statistic\tablenotemark{a} &
  Sample &
  \begin{tabular}{c}Coordinate\\System\end{tabular} &
  \begin{tabular}{c}Moment\\Tested\end{tabular} &
  Value\tablenotemark{b}}
\tablecaption{Dipole and quadrupole statistics for non-triggered GRBs
\label{tbl:isostat}}
\startdata
$\langle \cos \theta\ \rangle$ & Non-triggered & Galactic & Dipole & $-0.0029 \pm 0.0195$ \nl
$\langle \sin^2 b - \frac{1}{3} \rangle$ & Non-triggered & Galactic &
Quadrupole & $-0.0015 \pm 0.0101 $ \nl
$\langle \sin \delta \rangle$ & Non-triggered & Equatorial & Dipole & $0.0087 \pm 0.0195$ \nl
$\langle \sin^2 \delta - \frac{1}{3} \rangle$ & Non-triggered & Equatorial &
Quadrupole & $ 0.0191 \pm 0.0101 $ \nl
$\langle \cos \theta\ \rangle$ & Combined & Galactic & Dipole & $-0.0063 \pm 0.0121$ \nl
$\langle \sin^2 b - \frac{1}{3} \rangle$ & Combined & Galactic &
Quadrupole & $0.0018 \pm 0.0063 $ \nl
$\langle \sin \delta \rangle$ & Combined & Equatorial & Dipole & $0.0077 \pm 0.0121$ \nl
$\langle \sin^2 \delta - \frac{1}{3} \rangle$ & Combined & Equatorial &
Quadrupole & $ 0.0085 \pm 0.0063 $ \nl
$\langle \cos \theta\ \rangle$ & $P < 0.3$ ph cm$^{-2}$ s$^{-1}$ & Galactic & Dipole & $-0.0352 \pm 0.0223$ \nl
$\langle \sin^2 b - \frac{1}{3} \rangle$ & $P < 0.3$ ph cm$^{-2}$ s$^{-1}$ & Galactic &
Quadrupole & $0.0011 \pm 0.0115 $ \nl
$\langle \sin \delta \rangle$ & $P < 0.3$ ph cm$^{-2}$ s$^{-1}$ & Equatorial & Dipole & $0.0526 \pm 0.0223$ \nl
$\langle \sin^2 \delta - \frac{1}{3} \rangle$ & $P < 0.3$ ph cm$^{-2}$ s$^{-1}$ & Equatorial &
Quadrupole & $ 0.0293 \pm 0.0115 $ \nl
\enddata
\tablenotetext{a}{$\theta$ is the angle between the source direction
  and the Galactic center, $b$ is the Galactic latitude, and $\delta$ is
  the declination.}
\tablenotetext{b}{Moments are corrected for sky exposure.}
\normalsize
\end{deluxetable}

\begin{table}
  \dummytable\label{tbl:unktbl}
\end{table}

\begin{deluxetable}{lllr}
\small
\tablehead{Statistic\tablenotemark{a} &
  \begin{tabular}{c}Coordinate\\System\end{tabular} &
  \begin{tabular}{c}Moment\\Tested\end{tabular} &
  Value\tablenotemark{b}}
\tablecaption{Dipole and quadrupole statistics for low-energy
  ``unknown'' events
\label{tbl:unkiso}}
\startdata
$\langle \cos \theta\ \rangle$ & Galactic & Dipole & $-0.039 \pm 0.082$ \nl
$\langle \sin^2 b - \frac{1}{3} \rangle$ & Galactic &
Quadrupole & $0.015 \pm 0.043 $ \nl
$\langle \sin \delta \rangle$ & Equatorial & Dipole & $0.010 \pm 0.082$ \nl
$\langle \sin^2 \delta - \frac{1}{3} \rangle$ & Equatorial &
Quadrupole & $-0.002 \pm 0.043 $ \nl
\enddata
\tablenotetext{a}{$\theta$ is the angle between the source direction
  and the Galactic center, $b$ is the Galactic latitude, and $\delta$ is
  the declination.}
\tablenotetext{b}{Moments are corrected for sky exposure.}
\normalsize
\end{deluxetable}

\clearpage

\setcounter{figure}{0}

\begin{figure}
\plotone{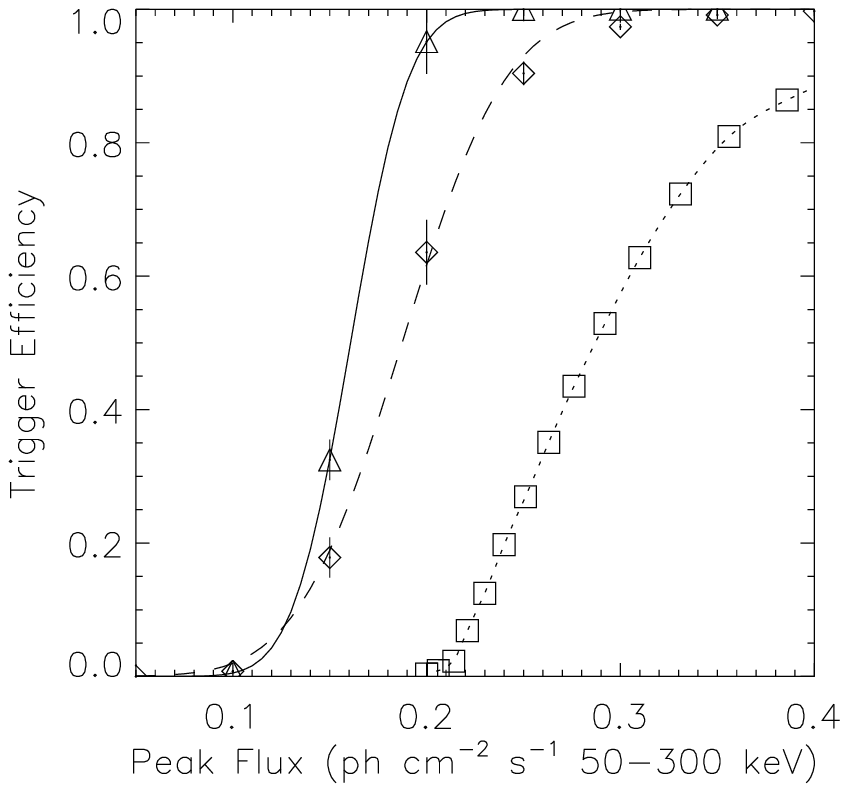}
\caption{~}
\end{figure}

\begin{figure}
\plotone{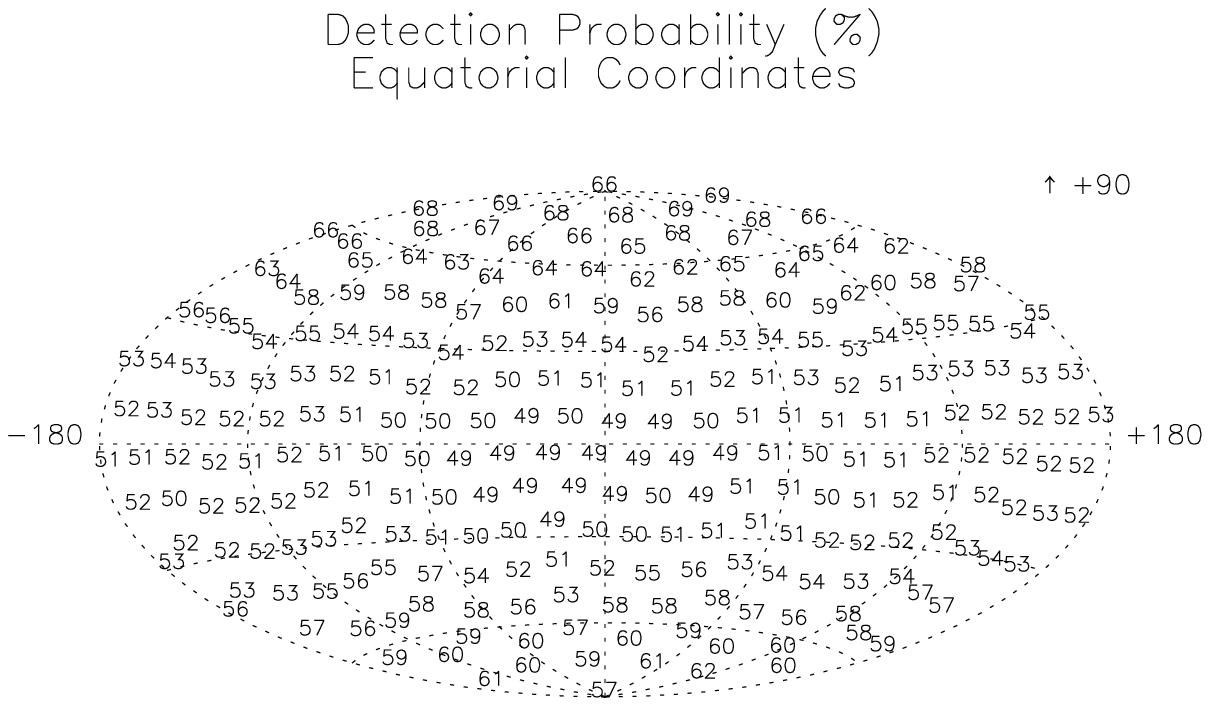}
\caption{~}
\end{figure}

\begin{figure}
\plotone{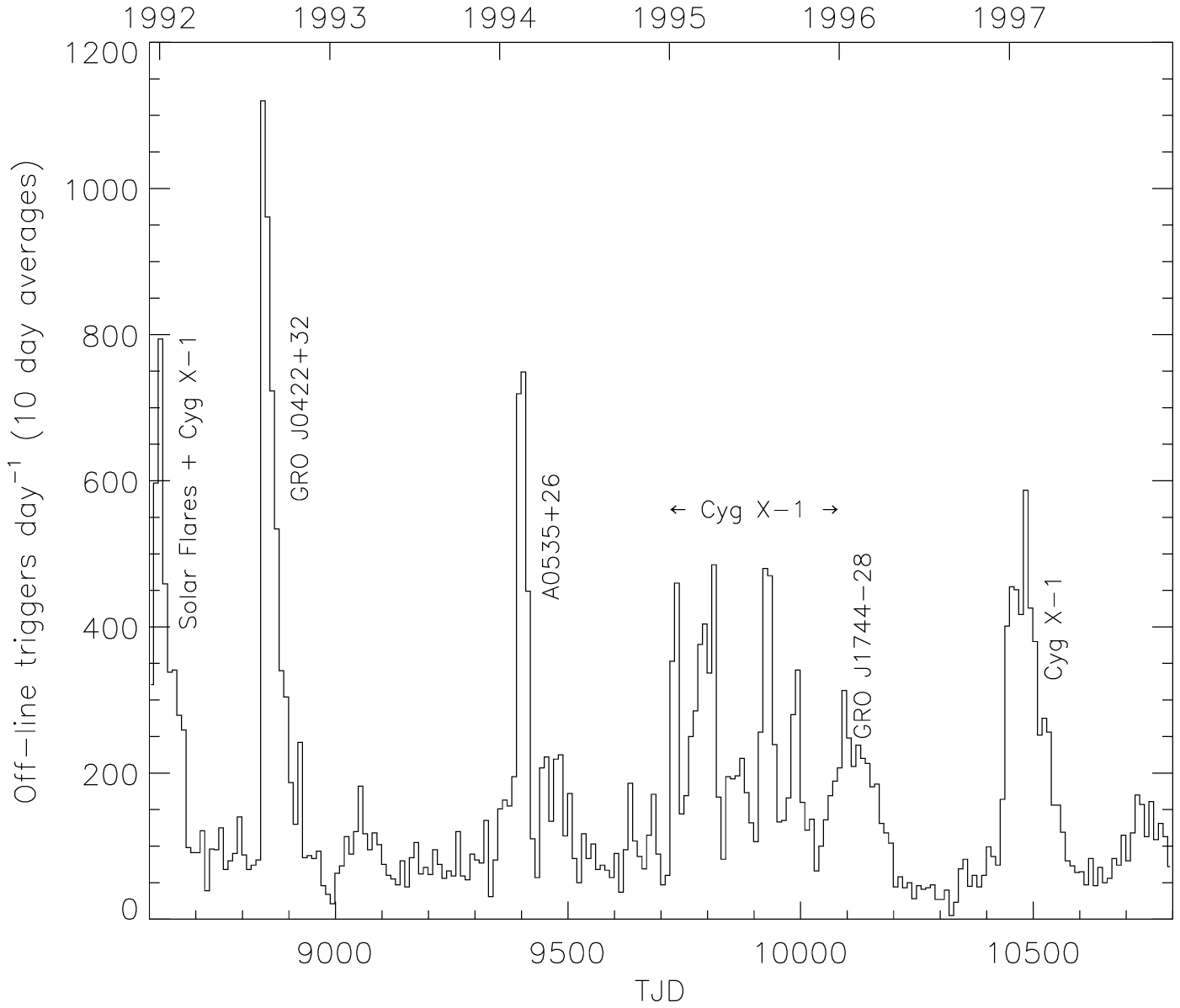}
\caption{~}
\end{figure}

\begin{figure}
\plotone{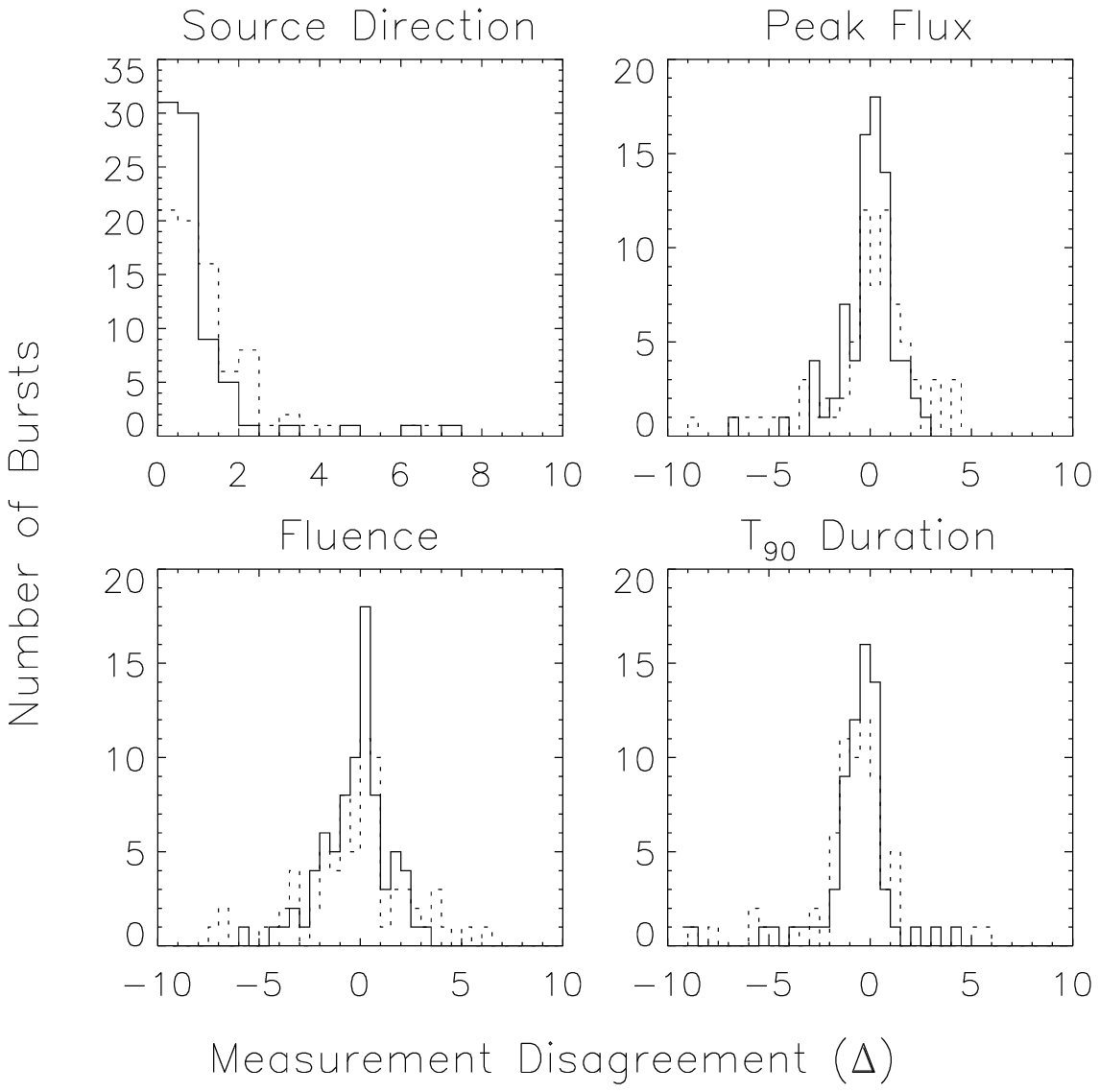}
\caption{~}
\end{figure}

\begin{figure}
\plotone{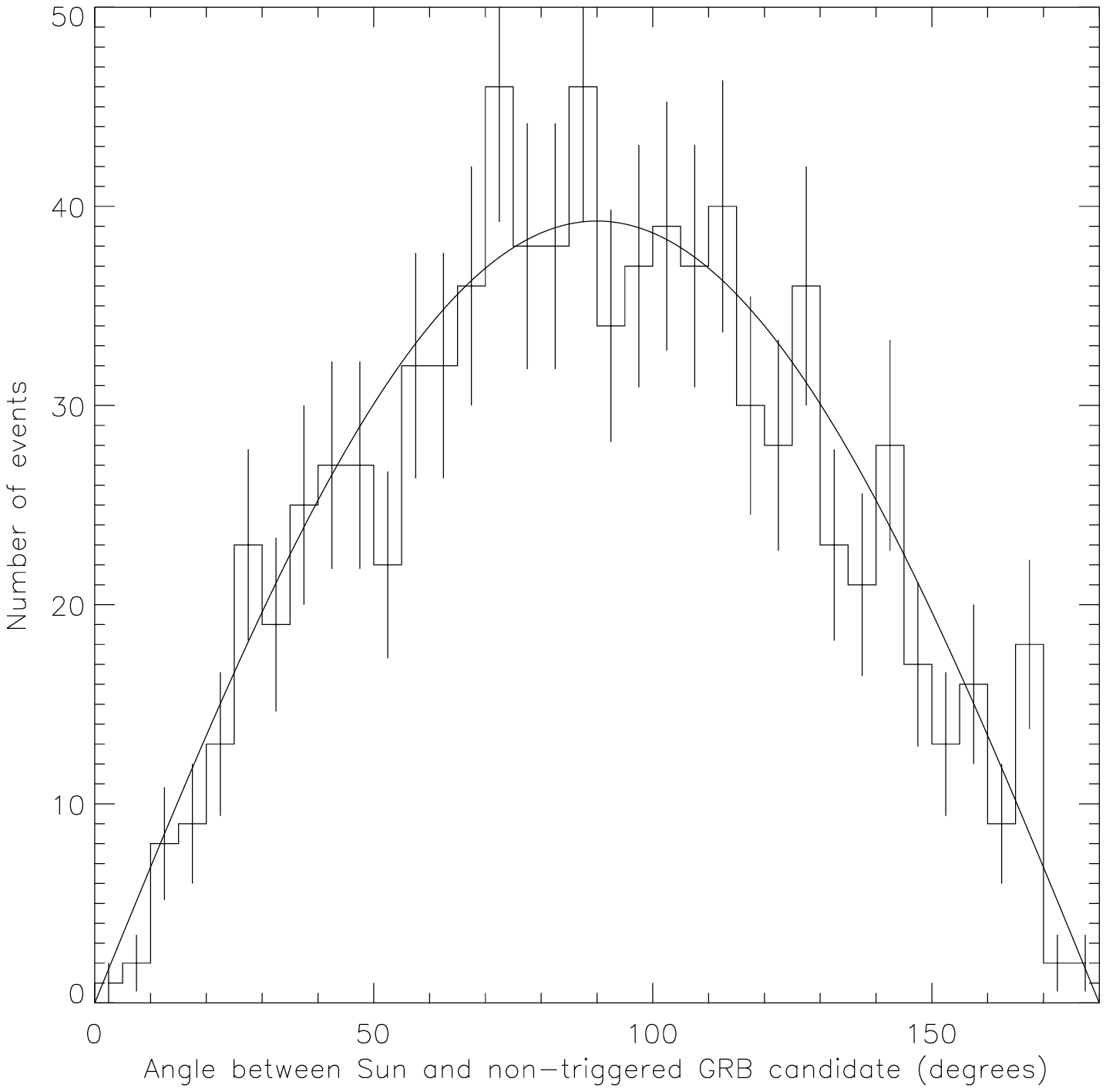}
\caption{~}
\end{figure}

\begin{figure}
\plotone{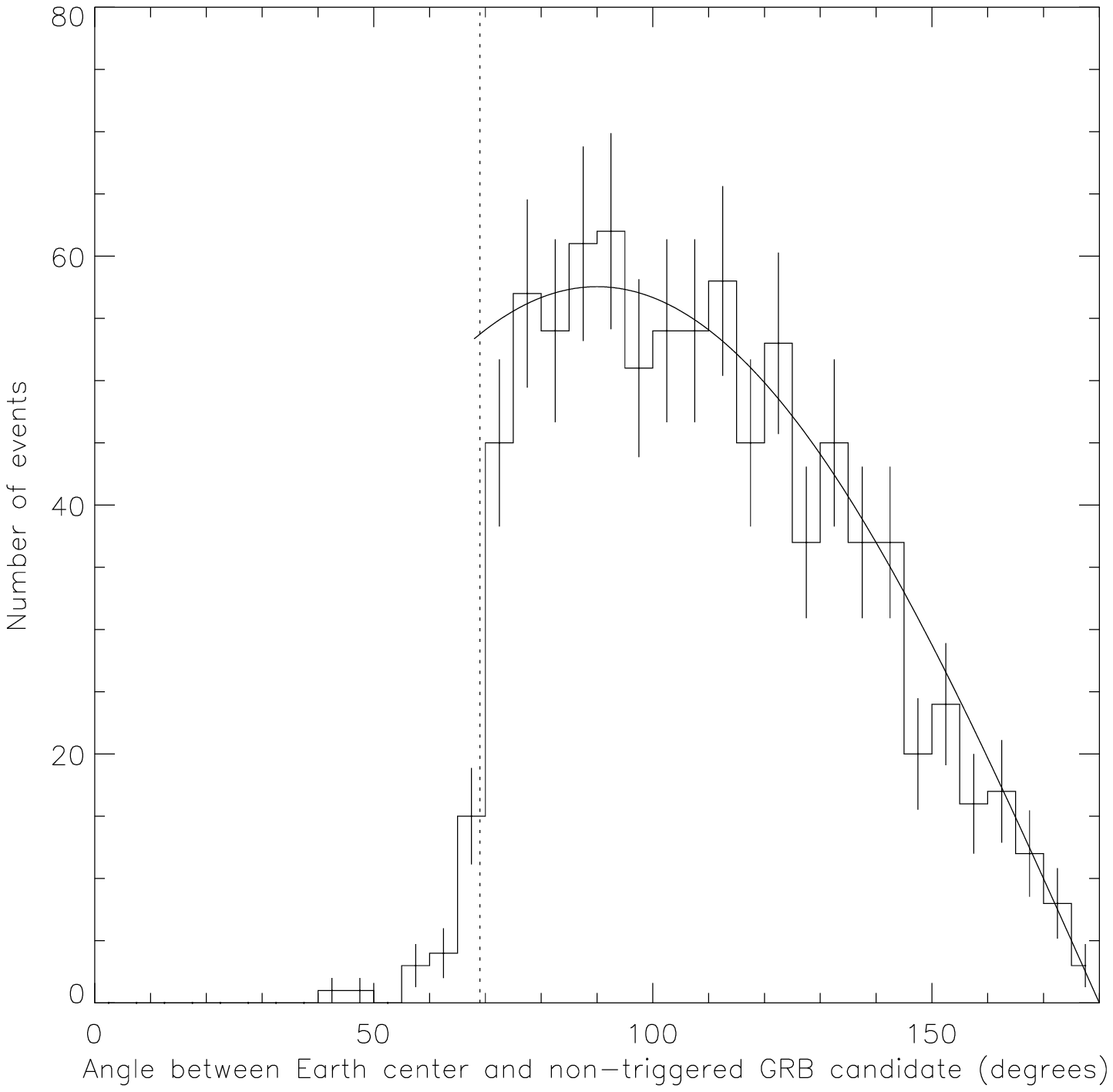}
\caption{~}
\end{figure}

\begin{figure}
\plotone{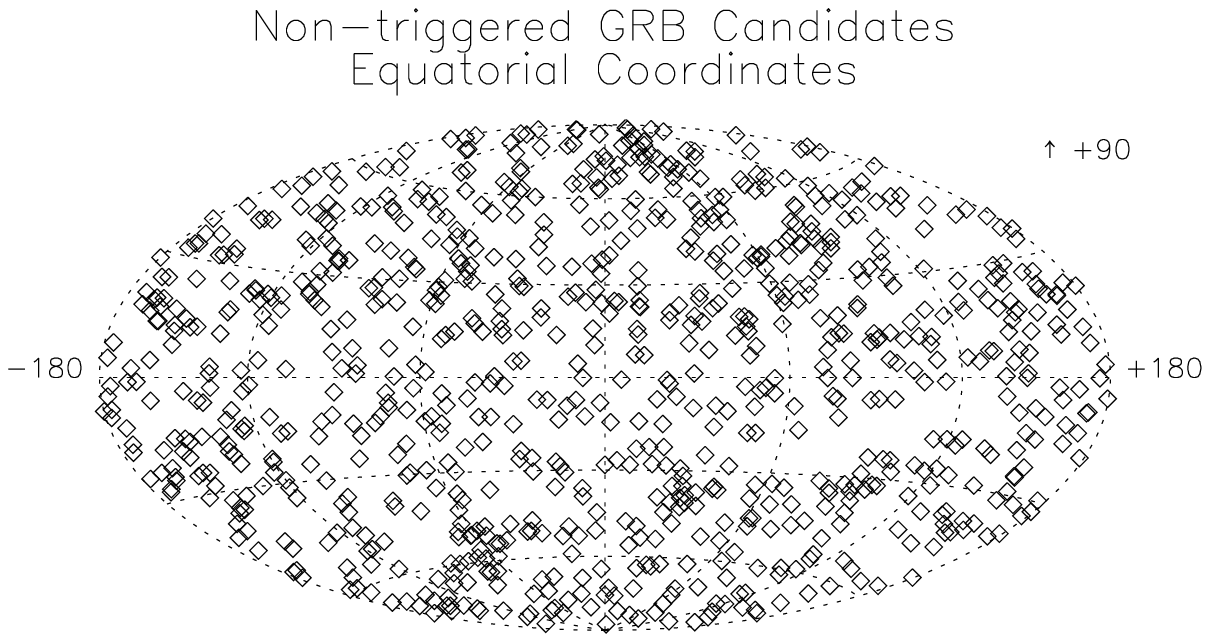}
\caption{~}
\end{figure}

\begin{figure}
\plotone{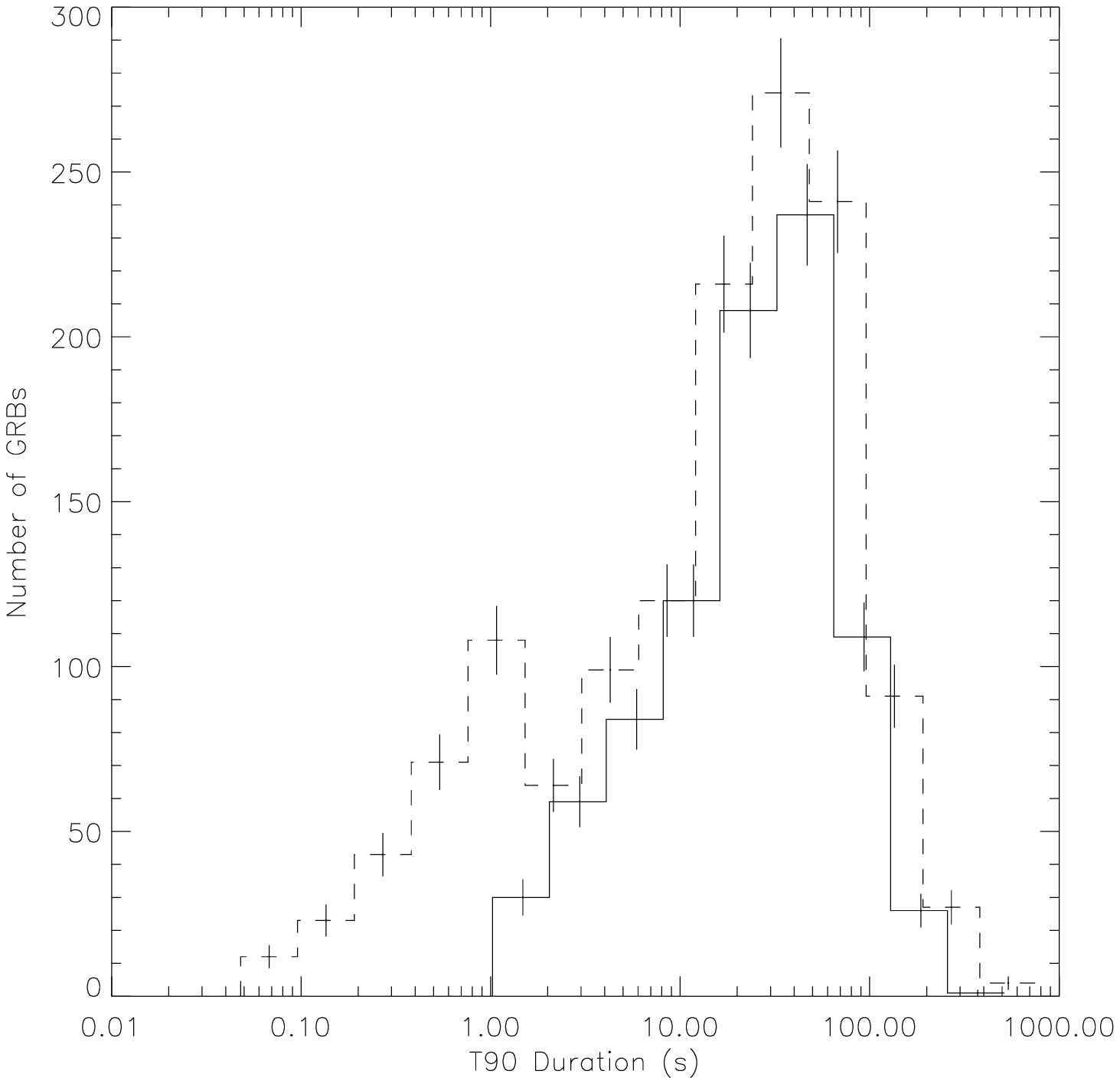}
\caption{~}
\end{figure}

\begin{figure}
\plotone{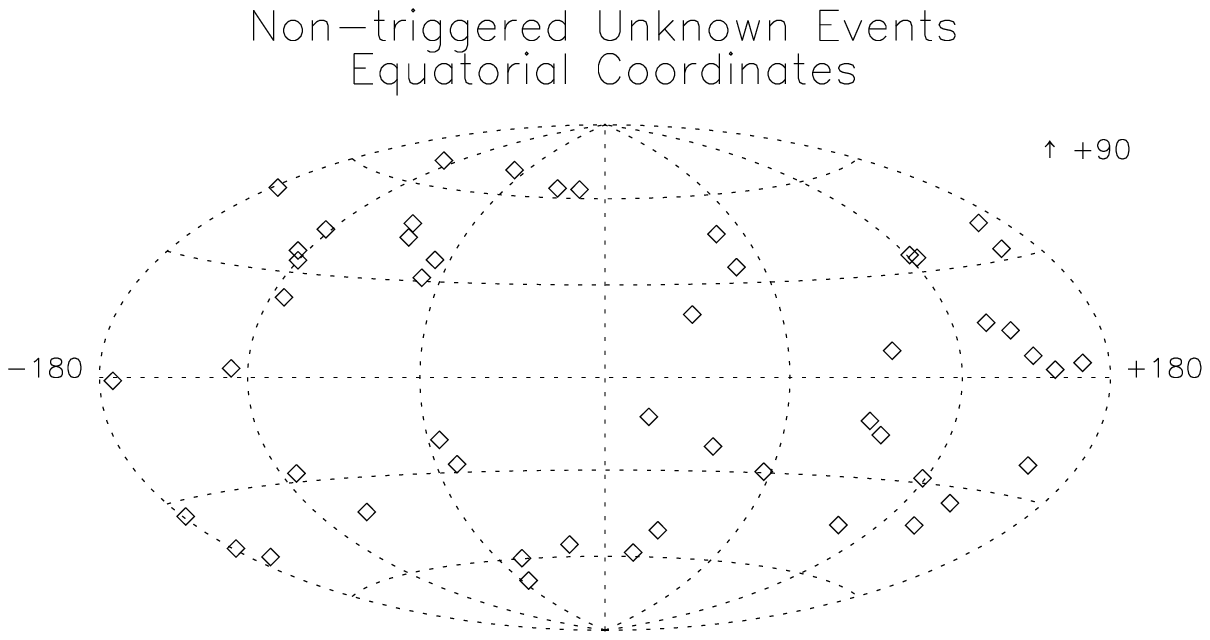}
\caption{~}
\end{figure}

\begin{figure}
\plotone{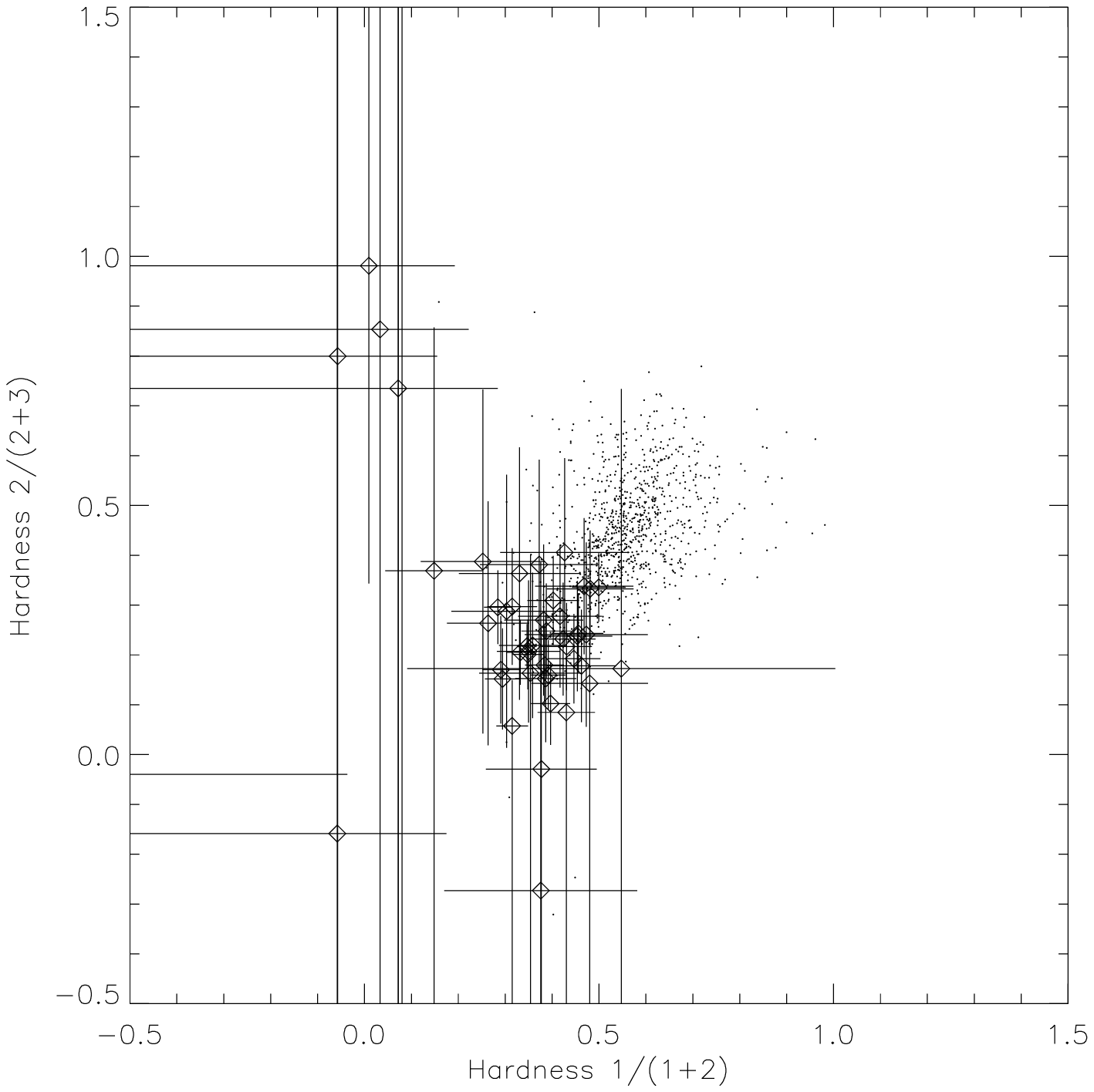}
\caption{~}
\end{figure}

\end{document}